\documentclass[format=sigconf]{acmart}
\pdfoutput=1
\newtheorem{property}{Property}[section]
\usepackage{algorithmic}
\usepackage{algorithm}
\DeclareMathOperator*{\argmax}{argmax}   
\usepackage{subcaption,siunitx,booktabs}
\usepackage{pgfplots}
\usepackage{tikz}
\pgfplotsset{compat=1.12}

\usetikzlibrary{plotmarks}
\usepgfplotslibrary{patchplots} 
\usetikzlibrary{patterns}
\usepackage{multicol}
\usepackage{multirow}
\usepackage{color,soul}
\newcommand{\fo}{flex-offer}
\usepackage{booktabs}
\renewcommand\footnotetextcopyrightpermission[1]{}
\usepackage[draft]{todonotes}
\usepackage{subcaption}

\newtheorem{assumption}{User Assumption}
\newcommand\blfootnote[1]{%
  \begingroup
  \renewcommand\thefootnote{}\footnote{#1}%
  \addtocounter{footnote}{-1}%
  \endgroup
}

\begin{document}

\title{Adaptive User-Oriented Direct Load-Control of Residential Flexible Devices}

\author{Davide Frazzetto}
\orcid{0000-0002-5809-8050}
\affiliation{%
  \institution{Aalborg University}
}
\email{davide@cs.aau.dk}

\author{Bijay Neupane}
\orcid{0000-0001-8559-2794}
\affiliation{%
  \institution{Aalborg University}
}
\email{bn21@cs.aau.dk}

\author{Torben Bach Pedersen}
\orcid{0000-0002-1615-777X}
\affiliation{%
  \institution{Aalborg University}
}
\email{tbp@cs.aau.dk}

\author{Thomas Dyhre Nielsen}
\orcid{0000-0002-4823-6341}
\affiliation{%
  \institution{Aalborg University}
}
\email{tdn@cs.aau.dk}

\begin{abstract}
Demand Response (DR) schemes are effective tools to maintain a dynamic balance in energy markets with higher integration of fluctuating renewable energy sources. 
DR schemes can be used to harness residential devices' flexibility and to utilize it to achieve social and financial objectives. 
However, existing DR schemes suffer from low user participation as they fail at taking into account the users' requirements. 
First, DR schemes are highly demanding for the users, as users need to provide direct information, e.g. via surveys, on their energy consumption preferences. 
Second, the user utility models based on these surveys are hard-coded and do not adapt over time. 
Third, the existing scheduling techniques require the users to input their energy requirements on a daily basis.
As an alternative, this paper proposes a DR scheme for user-oriented direct load-control of residential appliances operations.
Instead of relying on user surveys to evaluate the user utility, we propose an online data-driven approach for estimating user utility functions, purely based on available load consumption data, that adaptively models the users'  preference over time.
Our scheme is based on a day-ahead scheduling technique that transparently prescribes the users with optimal device operation schedules that take into account both financial benefits and user-perceived quality of service. To model day-ahead user energy demand and flexibility, we propose a probabilistic approach for generating flexibility models under uncertainty.
Results on both real-world and simulated datasets show that our DR scheme can provide significant financial benefits while preserving the user-perceived quality of service.
\end{abstract}

\maketitle

\section{Introduction}
\label{sec:intro}

The uncertainty in the power supply due to fluctuating Renewable Energy Sources (RES) has severe implications (financial and others) for energy market players. 
Traditional solutions such as the curtailment and the use of costly auxiliary services that market players utilize for compensating deviations between supply and demand lead to loss of revenue. 
\blfootnote{To appear in the Proceedings of the e-Energy 2018, ninth ACM International Conference on Future Energy Systems (ACM e-Energy 2018) }
Several smart grid projects have already addressed this problem, aiming at mitigating the effects of imbalances due to the integration of RES, by proposing Demand Response (DR) strategies for load shifting control \cite{AGP2013, Kobus2015, WL2015}. 
In this regard, the concept of utilizing the flexibility in household energy demand to dynamically balance the available RES with the energy load is one of the most promising. 
The goal is to capture the shiftable portion of energy as flexible consumption descriptions, so-called \fo s (proposed in MIRABEL\cite{mirabel, Boehm2012, LS2015, LS2016, EV2016, Emmanouil2015}), at the individual device level to obtain the highest resolution for flexibility control \cite{neupane2017generation}. 
In \cite{neupane2014towards, neupane2017generation, neupane2015evaluating} it has been shown that residential household devices such as dishwashers, washing machines, refrigerators, electric heating, heat pumps, and electric vehicles present opportunities for high flexibility at low (user) cost.
In this paper we focus on wet devices, such as dishwashers and washing machines, as they alone account for approximately 30$\%$ of the total energy consumption in residential households \cite{neupane2014towards}.

Recent studies have already researched how to exploit the flexibility potential of household devices for different objectives, e.g., increase smart grid capacity, maximize use of RES, or decrease energy costs, via Demand Side Management (DSM) programs based on user incentive design or dynamic pricing\cite{deng2015survey}. 
The former explores the possibility to shift the time and amount of load consumption to reduce the load at peak hours \cite{klaassen2016load, paterakis2015optimal, adika2014autonomous, Kobus2015}. 
In the latter \cite{li2011optimal, ma2016residential, astaneh2015novel}, instead of directly controlling the devices, the proposed programs indirectly encourage the users to change their energy usage schedules according to a dynamic pricing mechanism, as users are prone to use less electricity when the prices are higher.  
Most of the proposed demand shifting based DR schemes require the availability of accurate real-time and predicted future information on energy requirements, e.g., next 24 hours estimated energy needs, maximum device flexibility, device usage preferences/priorities, and manual device operation scheduling. 
This type of information has to be provided directly by the user via surveys or smart applications and has been shown to be too taxing for most users. 
In a user survey regarding home heating automation \cite{jensen2016heatdial} it was reported that \textit{"[Users] do not want to sit and regulate the heating every night"}, and in \cite{yang2014making} the authors discuss that because of time consuming user interaction, users do not use manual heating schedules and fail to reassess existing control patterns. 
As a consequence, it has been reported that users tend to drop out of more complex DSM programs in favor of  default non-automated flat rate energy plans--- a tendency denoted as \textit{user response fatigue} \cite{kim2011common}.  
Since energy consumption has historically been a passive purchase routine, the more interaction the user is required to perform, the higher the chances the users will abandon DSM programs. 

Regarding the estimation of day-ahead device-level load, there are a number of ongoing research projects looking at the possibility of utilizing device-level load data for dynamic DR \cite{Gottwalt2016, Develder2016, Yu2016}; even so, an efficient method for predicting device usage patterns and the associated energy demand is still missing as the unpredictability of the user behavior creates challenges in achieving higher accuracy for device-level demand forecasting.
 Further, predictive models for estimating flexibility in device usage, in order to devise an effective schedule for demand deferral, have been only partially addressed or still remains unexplored. 
In \cite{RA2014} the authors narrow the scope to predicting the deactivation times of currently operating devices.  This approach conflicts with the prediction horizon requirement of flexibility based DR, which requires beforehand flexibility information for efficient scheduling of supply and demand. 
Flexibility analysis based on the predicted demand has already been the focus of earlier work \cite{7799295, SM2013}.
However, the proposed models aim at predicting aggregated household demands, rather than individual devices.  Again, we emphasize that identification of device-level flexibility is paramount both to provide efficient load shifting DR schemes and to understand the end-user device usage behavior.

With respect to modeling and evaluation of user flexibility and scheduling of device flexible demand, we argue that direct load-control strategies should consider the user perceived loss of quality of service induced by a forced shifting in time of a device operation. 
Most of the existing research \cite{vardakas2015survey, deng2015survey, ma2016residential, tsui2012demand, setlhaolo2014optimal, samadi2012advanced, fahrioglu2001using, adika2014autonomous}  has focused on hard-coded approaches, independent from the specific user-device behavior.
Furthermore, current research has solely focused on modeling the level of satisfaction obtained by the user as a function of the \textit{amount} of energy consumption, rather than the \textit{time} of consumption, which is necessary to apply time-based scheduling techniques. 
In \cite{rikkeheat} the authors perform a qualitative user study that analyzes the wet-devices' usage behavior under a load shifting program. The study concludes that the users were not only able to shift their consumption to obtain financial benefits, but were also very willing to adapt to the new conditions. Nevertheless, the study involves heavy user-machine interaction, with no consideration for the risk of user response fatigue.

As an alternative, in this paper we propose a data-driven model for online estimation of the user utility as a function of device usage patterns. 
In order to minimize the burden of user-machine interaction, we propose solutions for estimating day-ahead device-level load consumption, modeling user flexibility, evaluating user preference for device operation, and prescribing a demand schedule that satisfies the user requirements. The contribution of the paper can be summarized as:
\begin{enumerate}
\item We present data-driven models for estimating user utility which significantly reduce the requirement of demanding user interaction and the threat of user response fatigue.
\item We propose a novel user-oriented direct-load scheme for scheduling of predicted flexible demand that considers both social  and financial aspects of demand shifting.
\item We present a novel method for modeling device-level flexibility under uncertainty.
\item We perform experiments on both real-world and synthetic device load datasets, and on a real energy market dataset. 
\item The experimental results show that our technique yields to an optimal trade-off between financial benefits and user-perceived quality of service, with up to $32\%$ and $38\%$ savings in the Spot and Regulation Market.
\end{enumerate}

The remainder of the paper is organized as follows. Section \ref{sec:prflschedule} discusses the flexibility concept and the proposed DR scheme for scheduling of flexible demand. Section \ref{sec:predictionModels} presents our approach for device-level activity prediction and how to model flexibility and uncertainty as probabilistic \fo s. Section \ref{sec:user_comfort} describes adaptive user flexibility estimation model. Section \ref{sec:pamodels} presents our scheme for user-oriented flexible demand scheduling. Section \ref{sec:experiments} presents experimental results and analysis. Finally, Section \ref{sec:conclusions} concludes the paper and provides directions for future work.
\section{Direct Load-Control of Flexible Devices}
\label{sec:prflschedule}
A DR scheme for direct-load control of flexible household devices involves multiple phases. Here we describe the workflow of our proposed DR scheme, and the modeling of flexibility into \fo s.

\subsection{DR Scheme Workflow}
\label{sec:dr_workflow}

Our proposed DR scheme focuses on \textit{day-ahead} demand deferral of flexible high-demand wet-devices. We choose a day ahead approach as it gives a better opportunity to avoid inbalances in the energy market \citep{deng2015survey}.
Starting with a practical example of our proposed scheme we consider the scheduling of a dishwasher $\mathcal{D}$, and through the example we will introduce the major concepts of our proposed scheme that will be further examined in the next sections.

Daily, we \textit{predict} which devices will be operated during the next day, the amount of demand needed by the operation, and the time in which this demand will be generated. 
The dishwasher $\mathcal{D}$ is predicted to operate at hour $h$ of day $d$. In this case, its \textit{flexibility} is extracted and a \textit{schedule} for deferring its time of operation (within the defined flexibility interval) is produced such that it fulfills market and user \textit{requirements}. 
Now, the user is informed directly via a mobile app, for example, of the maximum time for preparing the dishwasher ready to operate to comply with the proposed schedule. 
A user can always verify the proposed schedule using an application and at any time \textit{override} the schedule if so desired. In this case, the user receives no financial benefit.
If the user accepts the schedule and prepares it for the operation (fills it with dishes and marks it ready) no further interaction is required and the device operation takes place as scheduled unless interrupted by the user as described above. 
If the schedule is carried out successfully, the user receives a financial reward that depends on the amount of benefit the device flexibility provides to the energy market.

Our DR scheme is based on financial incentive-based scheduling of device operations while ensuring that users still have control over their devices. 
Compared to the related work \cite{klaassen2016load, paterakis2015optimal,samadi2012advanced, ma2016residential, WL2015} in which the users are required to directly provide information on their device and scheduling preferences, our proposed method requires fewer and simpler user interactions (e.g., only passive feedback from the device or smartphone via push notifications) and manual overriding of a proposed schedule.
It is important to note that complete elimination of user interaction is not only difficult to achieve but also undesirable as it has been shown that users want to feel in control of their devices and that feedback can help them understand the effects of home automation \cite{yang2014making, raptis2017aesthetic}. 
Nevertheless, reducing the burden of taxing user interaction can lower the risk of user response fatigue. 
In case of more traditional devices that cannot be automated, our proposed method can suggest the user optimal times of operation for a device. 
For example, in case of traditional dishwasher, the only required user interaction is to set the suggested delay for  the device's operation at the moment of filling it with dishes.


\subsection{Flexible Operation Modeling}
\label{sec:flex_modeling}

In our DR scheme, we consider a \textit{smart device} whose operations can be externally controlled. The process describing the operation of a device includes various steps. For example, if we consider the operation of a dishwasher, the process starts with a user preparing and consuming her meal, then loading the dishwasher with the used dishes, and finally activating the device to perform the cleaning task. 
Hence, we can abstract the usage of the device into the \textit{prepare} process, leading to the need of the device, e.g., using the dishes, the \textit{ready} process, started by the \textit{ready} action, e.g. loading the device and setting it ready to operate, and the \textit{activation} process, started by the \textit{activate} action, leading to the actual operation of the device. Our DR scheme is based on two assumptions on the relationship between these processes:

\begin{assumption}[\textbf{Device Usage Independence}]
The timestamps of the \emph{prepare} and \textit{ready} processes of a device are independent of the timestamp of performing an \emph{activate} action on the same device. 
\end{assumption}

This assumption captures a realistic user behavior of performing a task based on their requirements and preferences rather than based on external influence, i.e., users would not  change their daily routine to cope with the market-induced schedule. 
For example, users always prefer to prepare their meals irrespectively of the time for cleaning the dishes. Based on this assumption, we can consider the normal user behavior, i.e., the \textit{prepare} and \textit{ready} processes, to not be affected by the scheduling of the device.


\begin{assumption}[\textbf{Schedule Delay}]
 A user would accept a device to be activated \textbf{after} it is ready for operation, but not \textbf{before}.
\end{assumption}

We assume that a user would not easily change their preference of performing the \textit{prepare} process, e.g., change meal time to comply with our schedule. Therefore, we have no control over the \textit{prepare} and \textit{ready} actions, and the scheduler can only influence the time of the \textit{activate} process without interfering with the user-preferred time of the other two processes.


\begin{figure}
\minipage{0.49\linewidth}
  \includegraphics[width=\linewidth]{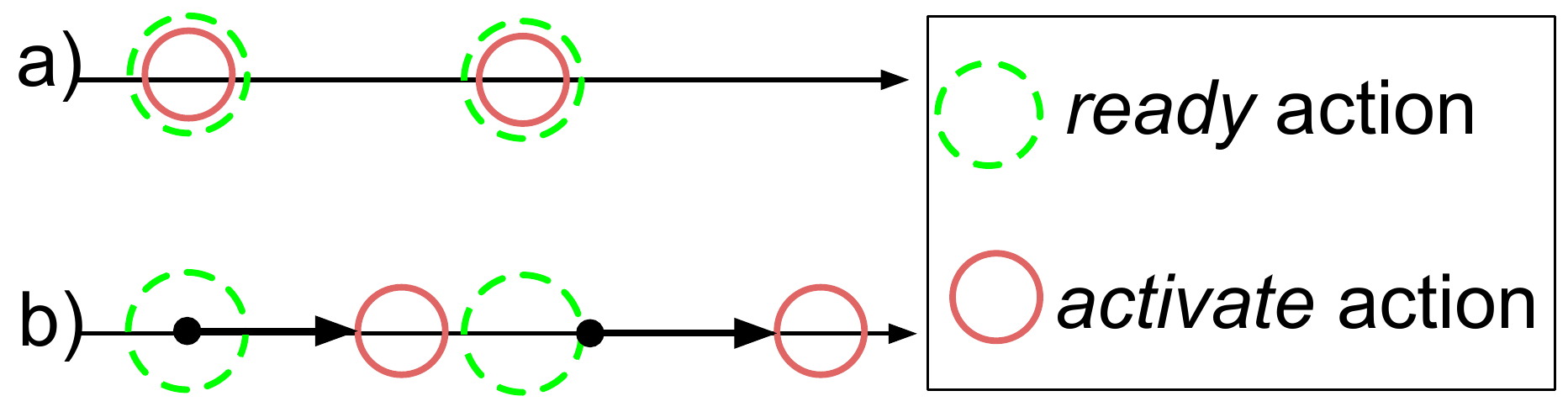}
  \caption{a) Non-flexible/non-controlled and b) flexible/controlled scenario.}
\label{fig:ready}
\endminipage\hfill
\minipage{0.49\linewidth}
  \includegraphics[width=\linewidth]{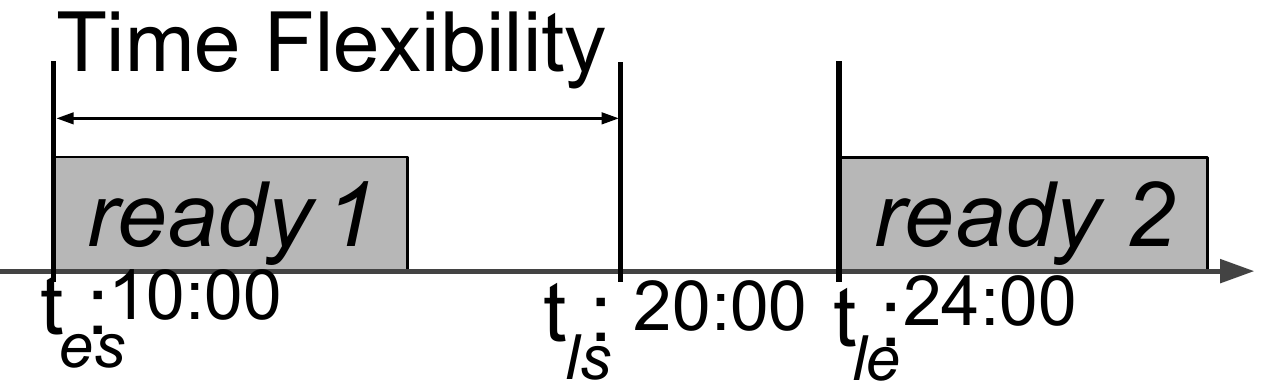}
 \caption{Flex-offer model for an activation/deactivation event of a device}
\label{fig:shiftinginterval}
\endminipage
\end{figure}

%

Figure \ref{fig:ready} shows a comparison between the operation of a device in a non-flexible/non-scheduled scenario (a), and in our proposed DR scheme (b).
On the one hand, in (a) each pair of \textit{ready} and \textit{activate} actions coincide, e.g., when a user loads and immediately starts a dishwasher, as the device provides no flexibility.
On the other hand, in (b), an \textit{activate} action can be delayed within its flexibility interval, i.e., as long as it does not interfere with the \textit{prepare} process that triggers the need of the device.

Our DR scheme relies on the availability of the timestamps of \textit{ready} actions. In historical data, as the one we have used in this work, \textit{ready} actions are represented by the timestamps of the device activation (assuming non-direct load-control) and in an online setup, the \textit{ready} actions represent the timestamps when user loads and marks ready their devices following normal behavior.
Since the \textit{prepare} process is user specific and may occur anytime between two \textit{ready} actions, we will only consider \textit{ready} and \textit{activate} actions for the proposed flexibility-based DR scheme.

\subsection{Flexibility as Flex-Offers}
\label{sec:flexibility}
We define the flexibility as the potential to amend the energy profile and demand deferral potential of a device, and we represent this flexibility as a so-called \textit{\fo}.   
The demand deferral potential, i.e. \textit{time flexibility}, represents the range between the \emph{earliest start time} ($t_{es}$) and the \emph{latest start time} ($t_{ls}$) for an operation (\textit{activate} action) of a device. 
For example, if we consider a single usage of a dishwasher, the time flexibility represents the earliest and latest time the dishwasher can be activated in order to fulfill a user's task. 
In more general terms, a \fo\ can define the boundaries within which future energy demands from a device can be scheduled.

\begin{definition}
\emph{A \fo\ $f$ is a tuple $\mathit{f = ([t_{es},t_{ls}], \rho)}$, where [$\mathit{t_{es}}\mathit{t_{ls}}$] is the time interval during which the \textit{activate} action can be shifted, $\rho = \mathit{\left \langle s_1,\dots, s_{d} \right \rangle}$ is the energy profile for the activation, $s_i =\mathit{[\mathit{e_{min}, e_{max}}]}$ is a continuous range defined by the minimum $\mathit{e_{min}}$ and maximum $\mathit{e_{max}} $ energy boundaries, and $\mathit{d}$ is the number of slices in $\rho$. }
\end{definition}

The latest end time of the device operation is calculated as $ t_{ls} = t_{ls} + d$. 
Although a \fo \ represents both time and amount flexibility,  the paper mainly focuses only on the time dimension of \fo s, as our target wet-devices do not allow for amount deferral, but only for time shifting, such that  $e_{min} = e_{max}$.
In Figure \ref{fig:shiftinginterval} we show an example of \fo\ with $t_{es} =  $ 10:00 and $t_{ls} =$ 20:00, in which we have a 10 hours of time flexibility range for rescheduling the device \textit{activate} event.

To conclude, a higher integration of RES into the grid system and electrification of user appliances (heat pump, electric vehicles) increases uncertainty in demand and supply, creating greater DR challenges for market players such as  Balance Responsible Parties (BRPs) and Distribution System Operators (DSOs). Flex-offers can be a valuable asset to the energy market players, where BRPs could utilize flex-offers to schedule demands that minimize their market deviations. Similarly, DSOs can use the \fo\ scheduling options to analyze the distribution of flexible loads within their grid system and assess the possibility of deferring expensive grid upgrades. The financial advantage obtained by the market players can then be shared with their customers. This way, end-users provide flexibilities to the market players, who can exploit it for financial benefits and share some portion of the benefits to the end-users. Further, One of the main objectives of modeling flexibility as a flex-offer is to have a generalized object capable of capturing and modeling flexibility from a variety of different devices.

\section{Prediction of Flex-Offers}
\label{sec:predictionModels}

In this section, we discuss the process of predicting device load consumption to generate the device's flexibility. We first describe the workflow of predicting device activity, and then we define an extension of \fo s that takes uncertainty into account.

\subsection{Device Activity Prediction}
\label{sec:dev_prediction}
The first step in the generation of flexibility is to predict the device's activity, in terms of future \textit{ready} actions. 
To do so, we first need to identify in which day the device will be operated, the time of the \textit{ready} action during that day, and the time until the next \textit{ready} action. 
The data we utilize for this process is load consumption time series at the device level. 
This type of data can be obtained directly from the device, e.g. with Smart Plugs, or by disaggregating Smart Meter data into individual device time series \cite{hosseini2017non}. 

First, we transform the load time series into an \textit{event} time series describing only the time of \textit{ready} actions. 
This is done by replacing the load consumption with the initial timestamp of each device operation. 
To identify device operations while maintaining the associated load/duration information, we abstract a device operation into a \textit{device signature}, representing a specific device operation in terms of both duration and energy consumption for time point of operation. 
A device signature is defined as:

\begin{equation}
\label{eq:signature}
\sigma = [e_1, e_2, \ldots, e_k ]
\end{equation}
where $e_i$ is the load demand per hour, and $k$ the (average) operation length in hours. Given the event time series, the activations are used to extract the average device operation duration and load demand per hour of operation. First, the energy demand per hour $e_i$ is estimated by averaging the hourly energy demand over all the activations in the event time series. Second, the operation length  $k$ is extracted as the ceiling of the average operation duration of each activation in the event time series.

\begin{figure}
\centering
\includegraphics[width=1\columnwidth]{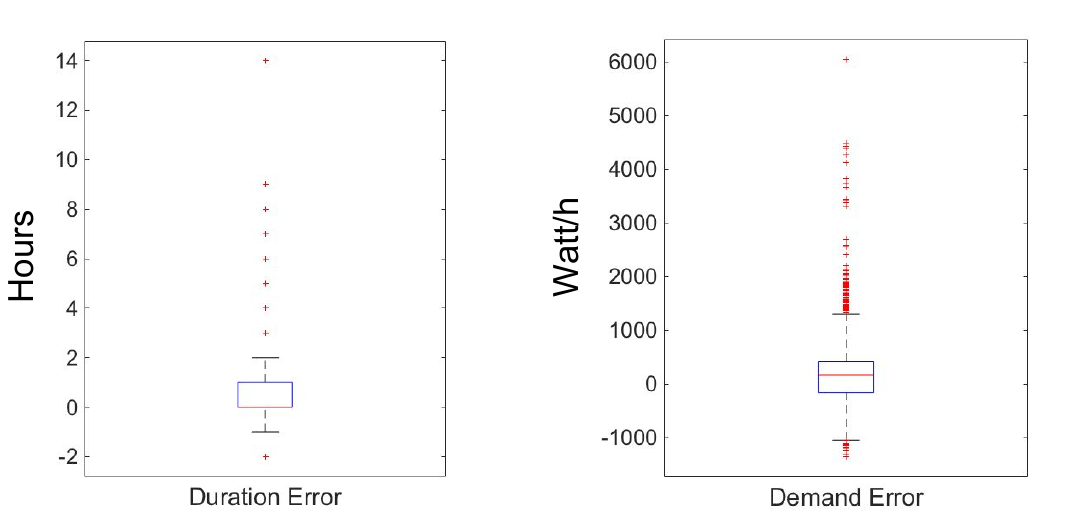}
\caption{Comparison of device signature and real activations in Hours (left) and energy consumption (right).}
\label{fig:signature_evaluation}
\end{figure}

In Figure \ref{fig:signature_evaluation} we show how the extracted device signatures resemble the actual device operations, in terms of the difference in duration between real activations and signature (left) and difference of total operation demand between real activations and signature (right) for all the devices in our datasets. 
It is possible to see that, given the short inter-quartile range in both plots, most the operations do not significantly differ from the device signature, allowing us to use the device signature as a mean to simplify the device load forecasting, without compromising the quality of the predictions. 
Although some outliers stretch beyond the upper and lower whiskers, they are less than $ 5\%$ of the total number of activations.

Second, we use the activation time series to perform a \textit{day-level} classification of whether the device will be operated during the next day. 
If a day is predicted to present an activation with a probability higher than $50\%$ (more information on the prediction models used in Section \ref{sec:dataset}), we also perform two  \textit{hour-level} predictions of the \textit{hour} of the next two consecutive \textit{ready} actions during that (or the next) day, in order to generate the earliest start time, first \textit{ready} action, and latest end time, second \textit{ready} action, of a \fo.

Let $T_{es}$ and $T_{le}$ be two random variables representing the time of the earliest start time and latest end time, respectively. Given that load time series are usually discrete, with the granularity of the sampling frequency or of the chosen aggregation level, e.g., 1 hour, we can also assume $T_{es}$ and $T_{le}$ to be discrete.

We first predict the probability of the time of the earliest start time $P(T_{es}|e)$, given an evidence set $e$, e.g., calendar information, day, month, week, etc. Then we predict

\begin{equation}
\label{eq:tle}
P(T_{le}, T_{es}|e)  = P(T_{le} | T_{es},e)P(T_{es}|e),
\end{equation}

as the probability of the time of the earliest start time and the latest end time, conditioned by the probability of $T_{es}$. 
Similarly, the probability $P(T_{le})$ can be defined as

\begin{equation}
P(T_{le}|e) = \sum_{T_{es}}P(T_{le} | T_{es}, e)P(T_{es}|e).
\end{equation}

\subsection{Probabilistic Flex-Offer}
\label{sec:prob_flex_offer}

A \fo\ can be used to model the device's flexibility in terms of earliest start time and latest end time, as described in Section \ref{sec:flexibility}. In our case, when modeling future flexibility from predicted device activity, earliest start time and latest start time are represented by the prediction over the two random variables $T_{es}$ and $T_{le}$. 
The definition of \fo\ given in Section \ref{sec:flexibility} can only represent the range between two individual time points, rather than random variables.
To overcome this limitation, we propose an extension to the standard \fo\, defined as a  \textit{Probabilistic \fo}.

\begin{definition}
\emph{A probabilistic \fo\ $f$ is a tuple $\mathit{f = \langle[T_{es}, T_{le}],}$ $\mathit{\rho\rangle}$, where $T_{es}$ and $T_{le}$ are two discrete random variables describing the earliest start time and the latest end time respectively, and $\rho$ is, as in a standard \fo\ , the energy profile of the activation.
$[T_{es}, T_{le}]$ defines a set of $|T_{es}| \times |T_{le}|$ possible flexibility intervals in which the activate action can be shifted, described by the tuple $\langle [t_{es}, t_{le} - |\rho|], P(T_{es} = t_{es}, T_{le} = t_{le}) \rangle$, where $t_{es} \in T_{es}$, $t_{le} \in T_{le}$, $|\rho|$ is the length of the operation, and $P(T_{es} = t_{es}, T_{le} = t_{le})$ is the interval probability defined in Eq. \ref{eq:tle}.}
\end{definition}

A probabilistic \fo\ considers not just a single flexibility interval between the earliest and latest start times, but all the intervals $[t_{es}, t_{ls}]$, with $t_{ls} = t_{le} - |\rho|$ between the time points described by the distributions over earliest start time and latest end time.
 Each of these intervals is also associated with the probability of such an interval to be correct, i.e. to accurately describe the actual interval between two consecutive device \textit{ready} actions.

\begin{figure}
\centering
\includegraphics[width=0.8\columnwidth]{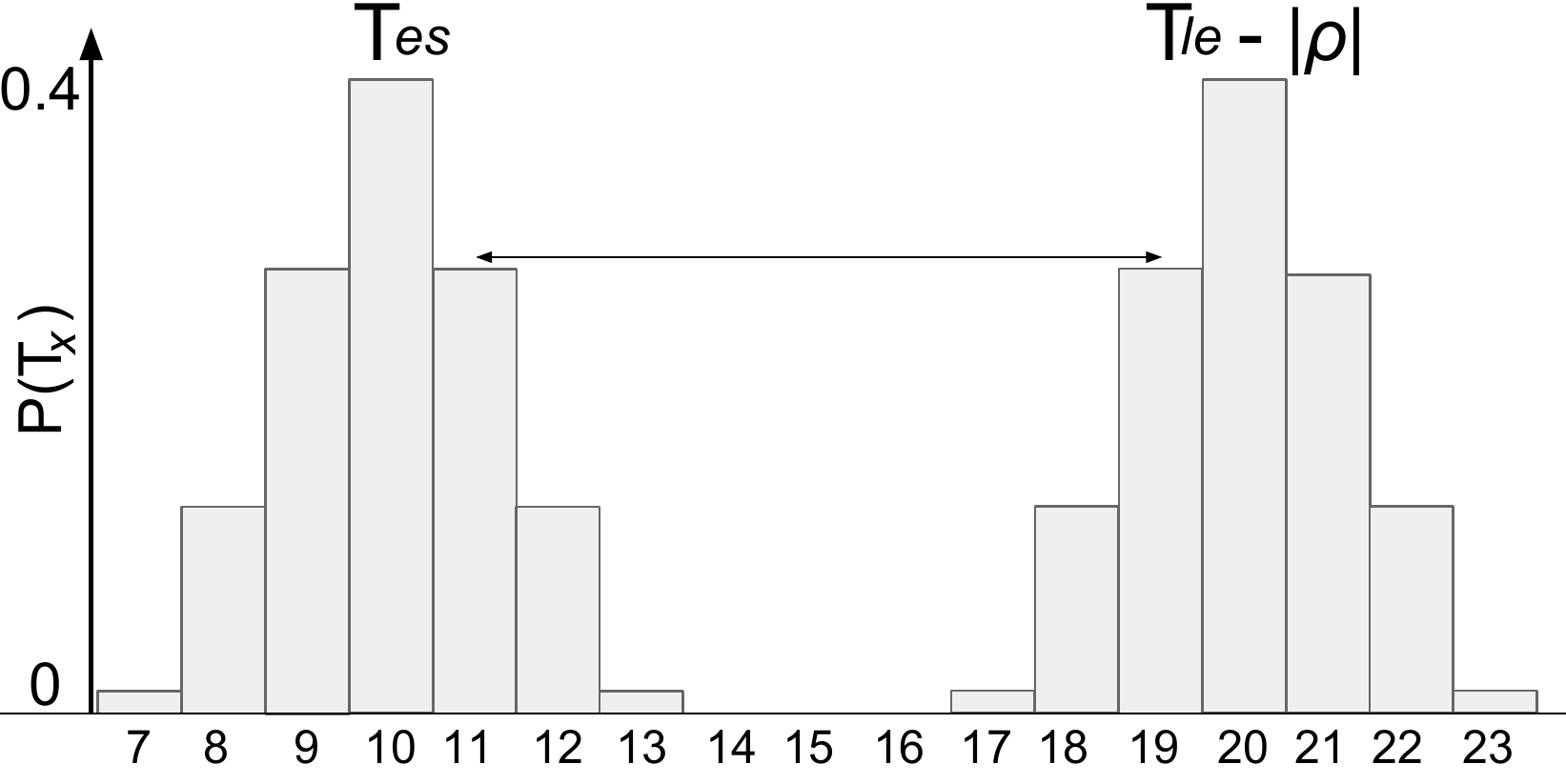}
\caption{Probabilistic \fo, the example shows one of the possible intervals between $T_{es}$ and $T_{ls} = T_{le} - |\rho|$, in this case  [11,19].}
\label{fig:prob_flex_offer}
\end{figure}

Figure \ref{fig:prob_flex_offer} shows an example  of the result of the prediction process for $P(T_{es}|e), P(T_{le}|e)$, showing the (truncated normal) probability distributions, with a granularity of 1 hour, of the random variables over the time ranges [7,13] and [17,23]. One of the possible flexibility intervals is [11,19]. In Section \ref{sec:pamodels} we will describe how to make use of probabilistic \fo s in order to more accurately schedule flexible demand.

\section{Adaptive User Utility Modeling}
\label{sec:user_comfort}
In this section, we describe our proposed model for data-driven user utility modeling.  We describe how to model the acceptance of a user towards a schedule as \emph{user utility}, a combination of financial benefits and user flexibility. Then, we show how to adaptively estimate the user flexibility directly from device-level load consumption data.

\subsection{User Utility}
In our proposed DR scheme, device schedules need to be (at least implicitly) approved by the owners. If the proposed schedules do not fulfill the users' requirements, users can actively diverge from them, resulting in a loss of (financial) opportunities for all the involved parties. 
So far we have modeled flexibility as a \fo. However, the flexibility interval described in such a model is an \textit{over-estimation} of the actual user flexibility, as it only describes the time between which the device is needed ready to operate.  
As an example, a user might not need to utilize the dishwasher until the day after, but the result of the dishwasher operation, i.e. clean dishes, is probably needed before that time. Therefore, understanding and respecting the user flexibility beyond what modeled so far in a \fo\ becomes paramount.

Each user is assumed to take decisions independently from the other users, the energy market, or the smart grid requirements. Specifically, residential users may follow a purely self-interest based \textit{utility} in their energy consumption.
This utility can be modeled as a combination of \textit{financial interests}, e.g. lower energy prices, and \textit{device interests}, the value of the operation of a device at a certain time of the day \cite{samadi2012advanced}. 
Hence, each user may possess a different utility function, resulting from different combinations of these two factors. As an example, a user might be more interested in the financial interests than in the device interests, thus not being concerned about the time of operation of her dishwasher as long as she is obtaining the highest financial benefit from the device operation. Alternatively, the same user might have a different utility if the scheduled device is a washing machine. Therefore, user utility is a function of the specific user-device pair, the time of the schedule, and the financial interests resulting from such a schedule.

Following this intuition, we start by introducing an assumption of our proposed user utility model:

\begin{assumption}[\textbf{User Utility}]
Users are \textit{flexible} in regards to their devices being rescheduled in return for financial benefits,  as long as the schedule matches their preferred device behavior.  Their acceptance of the device schedule is positively correlated to their financial interests and negatively correlated to the amount of delay induced by the schedule.
\end{assumption}

Intuitively, a user would, all else being equal, maintain control over the usage of the  devices,  i.e.,  to  not  reschedule  the activations.
However, according to our user utility assumption, we can represent  the  scheduling of a device activation as a trade-off between financial interests, providing higher flexibility, and device interests, constraining to lower flexibility. 
We will, therefore, define the expected User Utility, for a user $u$ and a device $d$, whose operation $o$ starting at time $t_{es}$ has been scheduled to time $t$, as

\begin{equation}
\label{eq:exp_utility}
\begin{split}
\mathbb{E}[U_{u,d}(t| t_{es},t_{le}, A)] &= G(A = a, t, t_{es}) \cdot P(A = a | t, t_{es}, t_{le}) \\
& + G(A=r, t, t_{es}) \cdot P(A = r | t, t_{es}, t_{le})
\end{split}
\end{equation}

where $G(A = a, t, t_{es})$ is the financial benefit obtainable when the proposed schedule is accepted, $A=a$, by delaying the operation of $o$ from $t_{es}$ to $t$, while $G(A = r, t, t_{es})$ is the financial benefit that would have been obtained by the rejected schedule, $A=r$. $P(A | t, t_{es}, t_{le})$ is the probability that the user would either accept, $P(A=a)$, or reject, $P(A=r)$, a schedule, with regards to the delay $t - t_{es}$ in the interval $[t_{es}, t_{le}]$, and user available flexibility. 
We can see the financial benefit $G(A, t, t_{es})$ as the savings obtainable by the lower energy price of the operation starting at $t$, compared to the original start time $t_es$, $G(t, t_{es}) = Price(o, t) - Price(o, t_es)$

When the proposed schedule is rejected we assume that the earliest start time is kept as operation starting time, i.e. $t = t_{es}$ and the user receives no financial benefit, thus $G(A =r, t, t_{es}) = Price(o, t_{es}) - Price(o, t_{es}) = 0$. Consequently, $G(A=r,t, t_{es}) \cdot P(A = r | t, t_{es}, t_{le}) = 0$, and Eq. \ref{eq:exp_utility} simplifies to 

\begin{equation}
\label{eq:expected_utility_2}
\mathbb{E}[U_{u,d}(t | t_{es},t_{le}, A)] = G(A=a, t, t_{es}) \cdot P(A = a | t, t_{es}, t_{le})
\end{equation}

Here, $G(A, t, t_{es})$ depends on the financial benefits, that will be described in Section \ref{sec:financial}, while $P(A| t_{es}, t_{le} )$ depends on the acceptance of the user, which we model as \textit{User Flexibility}.

\subsection{User Flexibility}

User flexibility can be described as the degree of acceptance of a user for a given device schedule, or as the probability that the user does not preempt the schedule proposed to a device operation. The objective is to understand the user preference in order to maximize the probability $P(A = a | t, t_{es}, t_{le})$, and consequently the expected utility of Equation \ref{eq:expected_utility_2}. 
To do so, we resort to utilizing a model based on \textit{exponential distributions}, whose natural interpretation is such that they describe the probability distribution of the time intervals between events in a stochastic process in which the events, in our case the $ready$ actions, occur at a constant average rate. 

Let $T$ be a random variable describing the distance between two $ready$ events of a device $d$, and $t - t_{es}$ be the delay imposed on a device operation $o$  by a schedule. Let us assume $t_{es} \leq t\leq t_{le} - |\rho|$, then $P(A=a |t, t_{es}, t_{le})$ simplifies into $P(A=a |t, t_{es})$.

Therefore, 
\begin{equation}
\label{eq:prob_f}
P(A = r | t, t_{es}) = P(T \leq t - t_{es}),
\end{equation}
where $P(T \leq t - t_{es})$ is the cumulative distribution function of the probability that a user will  need the device \textit{ready} before the proposed time $t$.
Alternatively, 
\begin{equation}
\label{eq:u_flex_f}
P(A = a | t, t_{es}) = 1 - P(T \leq t - t_{es}),
\end{equation}
represents the probability that a user accepts the schedule delay.
This way, we can model user flexibility functions as the cumulative distribution function $P(A = a | t, t_{es})$, starting by defining intuitive properties that user flexibility functions must possess.

\begin{property}
\textit{User flexibility functions are not increasing}
\begin{equation}
\frac{\partial (1 - P(T \leq t - t_{es}))}{\partial t} \leq 0,
\end{equation}
\end{property}

representing the partial derivative of the cumulative distribution function with respect to the schedule delay $t$. It described how the larger the delay $t$ induced by schedule, the smaller the probability the user would accept such a delay.

\begin{property}
\textit{When the delay $t-t_{es}$ is zero, the user acceptance is maximum}
\begin{equation}
1 - P(T \leq t_{es} - t_{es}) > 1 - P(T \leq t' - t_{es}), \quad T \geq 0, t' > t_{es},
\end{equation}
\end{property}
meaning that $1 - P(T \leq 0) = 1$.

\begin{property}
\textit{The probability a user accepts a schedule when the delay $t-t_{es}$ becomes larger tends to zero}
\begin{equation}
\lim_{(t-t_{es}) \to \infty} 1 - P(T \leq t - t_{es}) = 0.
\end{equation}
\end{property}
Meaning that a user loses all the utility of the device if the activation is scheduled too far from the original activation time. In our case, the exponential distribution function possesses the 3 properties, confirming as a good choice for describing user flexibility.

\subsection{Adaptive Estimation of User Flexibility}
\label{sec:adaptivity}
The current methods for modeling user utility functions already described in Section \ref{sec:intro} rely on direct user interaction, such as surveys or smart applications, for the estimation of the user utility functions. 
However, direct information from the users is often unavailable, as surveys are expensive to collect on a large scale or burden the user experience with unwanted interaction overhead, as already shown by a large number of users dropping out from DSM programs due to user response fatigue \cite{kim2011common}.
Further, user behavior can be temporary and may change over time, invaliding survey data.

To address this issue, we propose a data-driven user flexibility model purely based on device-level load consumption data. The objective is to estimate the function $1 - P(T \leq t - t_{es})$ defined in Eq. \ref{eq:u_flex_f}, for which we first need to estimate the distribution of $T$, i.e. the distribution of time between two \textit{ready} events.

\begin{figure}
    \centering
    \includegraphics[width=0.65\columnwidth]{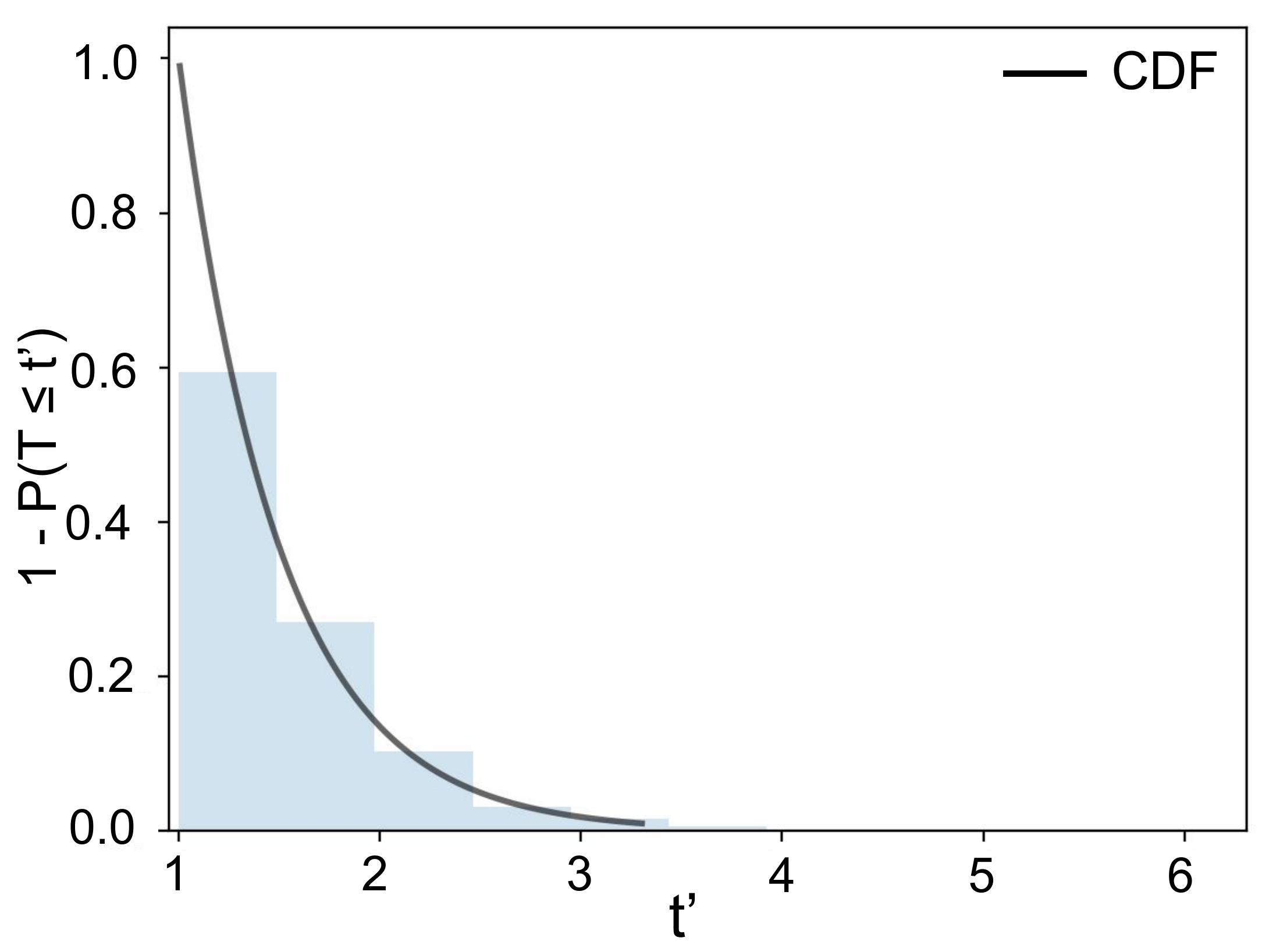}
    \caption{Inverse CDF---exponential distribution of the time intervals between \textit{ready} actions of a device.}
    \label{fig:cum_exp}
\end{figure}

Let $T$ be the random variable describing the distance (in hours) between two \textit{ready} actions, following an exponential distribution. Then the cumulative distribution function of $T$ can be defined as

\begin{equation}
\label{eq:pdf}
  F(t', \lambda)=\left\{
  \begin{array}{@{}ll@{}}
   1 -  \lambda e^{- \lambda t'}, & \text{if}\ t' \geq 0 \\
    0, & \text{otherwise}
  \end{array}\right.
\end{equation}

where $t' \in T$ is a time interval, and $\lambda$ is the \textit{rate parameter}, describing \textit{how often} two activations are separated by an interval of $t'$.
In our implementation, we generalize the $\lambda$ parameter to be a function of multiple factors, such as day of the week, month, and season. This way, we estimate a set of user flexibility functions rather than a single one, thus taking into account the possibly different behavior of the user in different time periods.
For our purposes, since the user flexibility is defined as  $1 - P(T \leq t - t_{es})$, we estimate the inverse of the cumulative distribution function from the distribution of $T$, as shown in Figure \ref{fig:cum_exp}.

To learn the parameter $\lambda$ of the exponential distribution we use a standard  Stochastic Gradient Descent method, by fitting the inverse of the cumulative distribution function of $T$ to the exponential distribution function. This approach can be applied both in an offline setting, when historical device data is available or in an online setting, where the user flexibility function updates every time a new user feedback is available.

For each iteration $i$ of the stochastic gradient descent algorithm, the parameter $\lambda$ is updated as in the following

\begin{equation}
\label{eq:gradient}
\lambda = \lambda- \mu \bigtriangledown Q_i(\lambda),
\end{equation}
where $\lambda$ is the parameter to be estimated by calculating the gradient of the objective function $Q_i(\lambda)$, and $\mu$ is the learning rate, i.e. how fast the parameter is changed in each iteration. $Q_i$ is associated with the i-\textit{th} observation. In an online setting, a new observation is given by the user feedback of either accepting or rejecting of the proposed schedule delay $t-t_{es}$. In case of acceptance, the observation is the scheduling delay $t-t_{es}$. Otherwise, the time of manual activation of the device by the user $t_m$, with $t_{es} \leq t_m < t$.

We choose a standard convex objective function, \textit{least-squares error}, that together with an adaptive learning rate, guarantees the convergence of the algorithm. 
Finally, to account for periods of times in which the user flexibility changes more drastically, e.g. different seasons, the $\lambda$ and $\mu$ parameters can be periodically re-initialized, e.g. every month, season, etc.

\section{Scheduling Device Operations}
\label{sec:pamodels}
The scheduling of flexible demand of a device is based on the User Utility, a combination of three factors. The first and the second are the financial gain that can be achieved in the spot and the regulation markets respectively by scheduling the device operation. The third is the User Flexibility, that we have analyzed in the previous Section. With the User Utility, we reflect the trade-off between maximizing financial gain and minimizing loss of user-perceived quality of service and thus propose a schedule that optimizes the combination of these factors. The next sections describe the financial factors in detail, and formally define the scheduling objective.

\subsection{Financial Model}
\label{sec:financial}

The adaptation of the flex-offer concept and demand scheduling depends on the financial value that flexibility can generate to all the involved parties, i.e., the savings for market players and rewards to the consumers. The market players can obtain saving by utilizing the demand flexibility to confront their challenges, such as to correct their deviations in the spot market or to avoid grid constraints, etc. However, some percentage of savings from the market players has to be transferred to the flexibility providers (consumers/prosumers) to maintain their commitment to the market. In this paper, we evaluate the savings from flexibility in two different markets. First, we evaluate the savings that can be obtained in the spot market by scheduling demands to a low energy price area. Second, we evaluate the savings that can be obtained by schedule demand in a way that reduces the energy traded on the regulation (intra-day) market. The decision on timestamp to shift the flexible demand depends on a combination of savings obtained in both the markets (discussed in the next sections) and also the acceptance of the new schedule (discussed in Section \ref{sec:user_comfort}).

\subsection{Spot Market Savings}
Spot market savings describes the total financial savings of energy demands and the corresponding \fo s at the spot market for the predicted device activation. To maximize this factor, a device operation is scheduled such that the cost of purchasing the energy required for a device operation is minimized. Let $\mathit{Spot = [\sigma^s(t_{es}), \dots, \sigma^s(t_{le})]}$ represent the hourly spot prices between the earliest start time and the latest end time.
The energy cost for each timestamp of the device operation is calculated as the product of energy demand and the respective spot price given by
\begin{equation}
\label{eq:spotp}
S(x) = \sum_{i=0}^{|\rho|-1}e_i \cdot \sigma^s(x + i),
\end{equation}
where $|\rho|$  is the duration of the device operation $o$ starting at a timestamp $x$ and $e_i$ is the demand for each operating time unit.
Therefore, the savings in spot market cost obtained by scheduling the device \textit{activate} action from $t_{es}$ to $t$  is given by

\begin{equation}
\label{eq:delta_s}
\begin{split}
\Delta S &= S(t) - S(t_{es}) \\
& = \sum_{i=0}^{|\rho|-1}e_i \cdot \sigma^s(t + i) - \sum_{i=0}^{|\rho|-1}e_i \cdot \sigma^s(t_{es} + i).
\end{split}
\end{equation}

\subsection{Regulation Market Savings}
Regulation market savings describes the total financial savings in the regulation market that can be achieved by scheduling the device's demand for the predicted device activation. 
To maximize this factor we schedule the operations such that the total energy traded in the regulation market is decreased, eventually reducing the financial loss from demand and supply imbalance.

{
    \def\OldComma{,}
    \catcode`\,=13
    \def,{%
      \ifmmode%
        \OldComma\discretionary{}{}{}%
      \else%
        \OldComma%
      \fi%
    }%
Let $V = \langle v_{u/d}(t_{es}), \dots,v_{u/d}(t_{le}) \rangle$ and $Reg=\langle \sigma^r_{u/d}(t_{es}),  \dots, \sigma^r_{u/d}(t_{le})\rangle$ represent regulation volumes and prices between EST and LET, where $\mathit{v_{u/d}(x)}$ denotes the nonzero elements of regulating volume, i.e., up- or down-regulation and $\mathit{\sigma^r_{u/d}(x)}$ is the predicted up-regulating power price $\sigma^r_u(x)$ in case of up-regulation and the predicted down-regulating power price $\sigma^r_d(x)$ in case of down-regulation.%
}
 For each hour in $V$, the loss due to the market imbalance is computed as a product of the regulation volume times the price difference between regulating and the spot price. Hence, the total regulation cost for an operation starting at $t_{es}$ is calculated as:
\begin{align}
R(t_{es}) = \sum_{i=1}^{|\rho|-1}v_{u/d}(t_{es} + i) \cdot |\sigma^r_{u/d}(t_{es} + i) - \sigma^s(t_{es} + i)|
\label{eqn_rcost}
\end{align}
where $\sigma^r_{u/d}(t_{es} +  i)$ and $\sigma^s(t_{es} + i)$ are regulation price and spot price at time $t_{es} + i$, respectively.

Given the regulation volume and the predicted device \textit{ready} event and associated demand, the market generates a device schedule, inducing a delay $t-t_{es}$, that minimizes the regulation volumes. 
Let the new expected regulation volumes be $$\bar{V} =  \{ \overline{v_{u/d}(t + i)}, \forall i \, | \, \overline{v_{u/d}(t + i)}  \leq v_{u/d}(t + i)\},$$ where the overbar denotes the change in regulation volume due to shifting of flexible demand. 

Finally, the savings in the regulation market cost obtained by scheduling the device \textit{activate} action from $t_{es}$ to $t$  is given by

\begin{equation}
\label{eq:delta_r}
\begin{split}
\Delta R &= R(t) - R(t_{es}) \\
&=  \sum_{i=1}^{|\rho|-1}\overline{v_{u/d}}(t + i) \cdot |\sigma^r_{u/d}(t + i) - \sigma^s(t + i)| \\
&- \sum_{i=1}^{|\rho|-1}v_{u/d}(t_{es} + i) \cdot |\sigma^r_{u/d}(t_{es} + i) - \sigma^s(t_{es} + i)|
\end{split}
\end{equation}

\subsection{User Utility based Scheduling}
\label{sec:scheduling}

The schedule of an operation $o$ described by a probabilistic \fo\ $f$ 
maps the initial time of the activate action $t_{es}$ to a new timestamp $t$, inducing a delay of $t-t_{es}$ in the device operation. The new timestamp is included between the probabilistic time flexibility interval $[T_{es}, T_{le}]$ of the \fo, remembering also that the latest start time is defined as $t_{ls} = t_{le} - |\rho|$.

Now, we can complete the definition of User Utility given in Eq. \ref{eq:expected_utility_2} by replacing $G(A,t, t_{es})$ with the sum of Spot and Regulation market savings, of Equations \ref{eq:delta_s} and \ref{eq:delta_r} respectively:

\begin{equation}
\label{eq:gain}
\mathbb{E}[U_{u,d}(t|t_{es}, t_{le},A)] = (\Delta S + \Delta R) \cdot P(A=a|t,t_{es}, t_{le}).
\end{equation}

However, the scheduling is also dependent on $P(t_{es}, t_{le}|e)$, i.e., that the flexibility interval $[t_{es}, t_{le}] \in [T_{es}, T_{le}]$ is correct.  
To take into account the uncertainty of the flexibility intervals, the objective function for the scheduling of an operation $o$ can be defined as

\begin{equation}
\label{eq:sch_objective}
\mathbb{E}[t] = \sum_{t_{es}, t_{le}} \mathbb{E}[U_{u,d}(t|t_{es}, t_{le}, A)] \cdot P(t_{es}, t_{le}|e)
\end{equation}
where $P(t_{es}, t_{le}|e)$ is the probability of the \fo\ interval $[t_{es}, t_{le}]$ defined in Eq. \ref{eq:tle}. We remind the reader that the $T_{es}$ and $T_{le}$ are \textit{discrete} random variables, thus Equation \ref{eq:sch_objective} calculates a summation over the discrete values in $[t_{es}, t_{le}]$ rather than an integration.

Therefore, the scheduling function for the operation $o$ selects the $t$ that maximizes the expected utility $\mathbb{E}[t]$ such that: 
\begin{equation}
\text{Sched(o)} =  \argmax_{t} \mathbb{E}[t].
\end{equation}

\section{Experiments}
\label{sec:experiments}
We performed a number of experiments to analyze the effectiveness of the proposed DR program. 
First, we show the evaluation of the \fo\ scheduling process. 
Second, we show how our proposed adaptive user flexibility estimation can help preserving user comfort.
Finally, we present the impact that the prediction of probabilistic \fo s has on the scheduling process.

\subsection{Experimental Setup Description}
\label{sec:dataset}
We utilize a real-world energy consumption time series at the device-level. The time series are associated with 11 devices (6 washing-machines, 5 dishwashers), collected from 11 different households, which are available from the open dataset Intrepid \cite{INTrEPID}. The logged granularity of the time series varies depending on the device, ranging from 1 minute to 1 hour, and spanning for 1 year.  In this paper, we consider scheduling of flexible demand at an hourly resolution. Hence we aggregate, by averaging, the time series accordingly. We pre-process the time series, and use them for the generation of probabilistic \fo s, as discussed in Sec \ref{sec:dev_prediction}.

The device datasets are highly unbalanced, with an average number of activations events ranging between $1\%$ and $7\%$ of the entire time series, while the average number of activations per day is approximately $0.7$. Also, load consumption time series at a device level present a high degree of \emph{irregularity}, due to the dynamic behavior of the user, and the device usage can vary due to external factors not included in the data.  Hence, to evaluate and compare the proposed scheduling methods for various user-specific attributes such as family size, occupation, etc., we utilize a synthetic dataset generated with the Genetx tool described in \cite{hoogsteen2016generation}. The dataset is composed of 26 devices---13 dishwashers and 13 washing machines, each belonging to a different category of households, e.g., single worker, a family with children, etc. Each of these devices is modeled with a dynamic amount of flexibility, depending on the time of operation and household category. Both the Intrepid and the Genetx datasets present similar device usage characteristics, with comparable time between ready events for the devices.
For evaluation purposes, we split both datasets into training ($80\%$) and test set ($20\%$). The forecasting model and the user flexibility model are trained on the training data. On the test set we perform a prequential evaluation, with each step being a device schedule proposal, with both the user flexibility models and the forecasting models updated at each step, to align to a possible real application scenario.

For the financial evaluation, we use the Danish energy market dataset for DK1 price zone obtained from Energinet\footnote{ www.energinet.dk/EN/El/Engrosmarked/Udtraek-afmarkedsdata/Sider/default.aspx}. To avoid dependency of the experiments on this specific energy market dataset, we perform multiple runs of each experiment, randomizing the energy dataset at each iteration, finally averaging the results.

For forecasting device activity, we used a Naive Bayesian model for the day-level prediction, and a Linear Regression model for the hour-level prediction. 
We chose these two generic models as they fit the two respective problems without specific tuning: a Naive Bayesian model for classification of which day presents activity and a Linear Regression model for the hour of activity prediction respectively.  As this paper does not focus directly on the challenge of Short-Term Load Forecasting at a device level, we decided to apply the two simple models as \textit{baselines} representatives of the existing forecasting models. Both models were trained on features directly extracted from event time series, the time such as day of the week, week, month, weekend/weekday, and season.

\subsection{Evaluation of Device Operation Scheduling}
\label{sec:prescr_results}

The acceptance of a schedule and the utility to the market players are affected by both the accuracy of the probabilistic \fo\ (at describing the time of device flexibility) and the modeling of the user flexibility. With regards to the user flexibility model, the stricter the model is, i.e., a model that always attempts to satisfy the user acceptance, the higher the acceptance. However, at the same time, a stricter model also provides less flexibility to the market, resulting in lower utility to both users and market players. Therefore, we emphasize that for an efficient implementation of flexibility-based markets, a scheduler should always perform a trade-off between user acceptance and financial benefits.

\begin{figure*}
\minipage{0.32\textwidth}
  \includegraphics[width=0.9\linewidth]{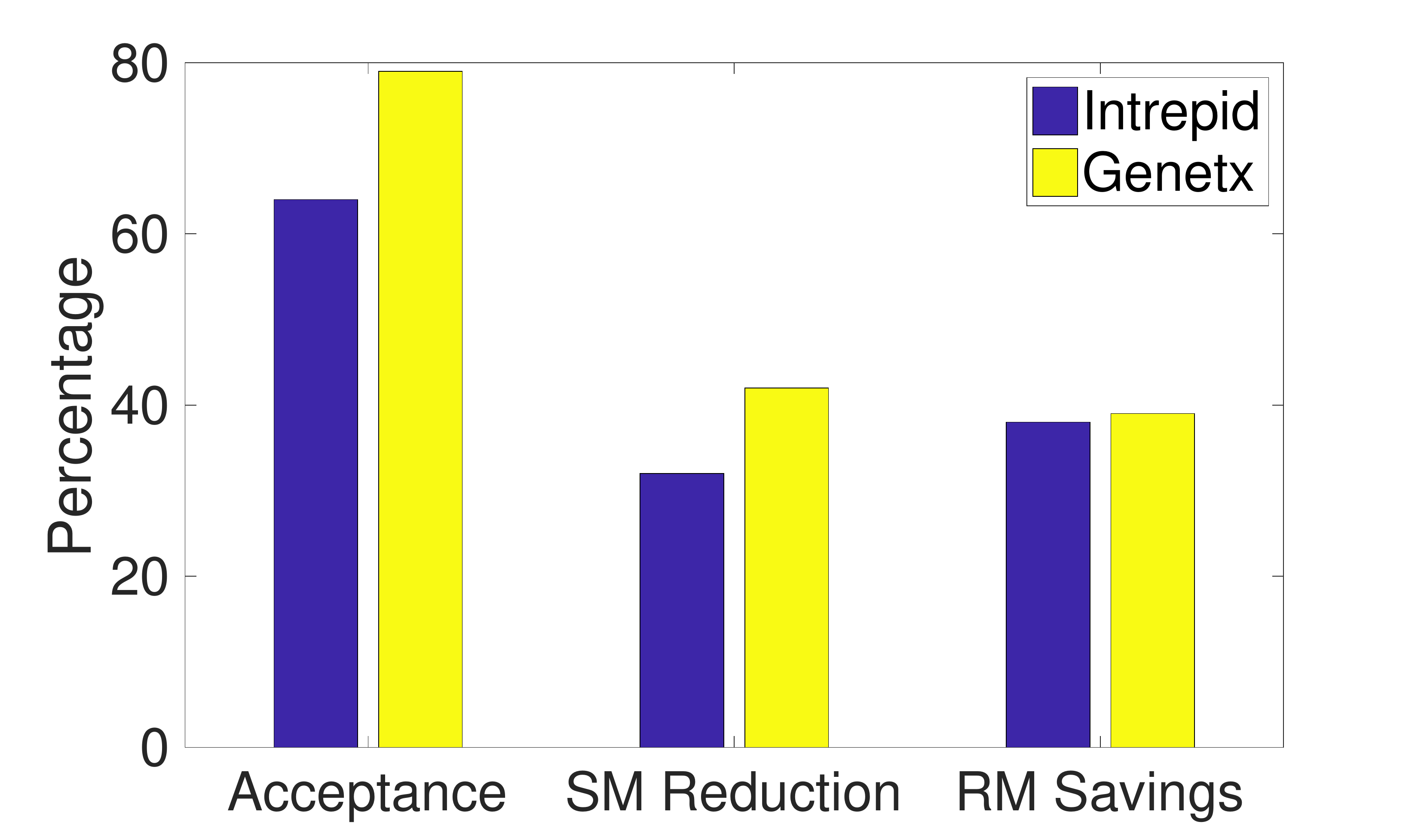}
  \caption{Scheduling results on Intrepid and Genetx datasets.}\label{fig:gain_summary}
\endminipage\hfill
\minipage{0.32\textwidth}
  \includegraphics[width=1.3\linewidth]{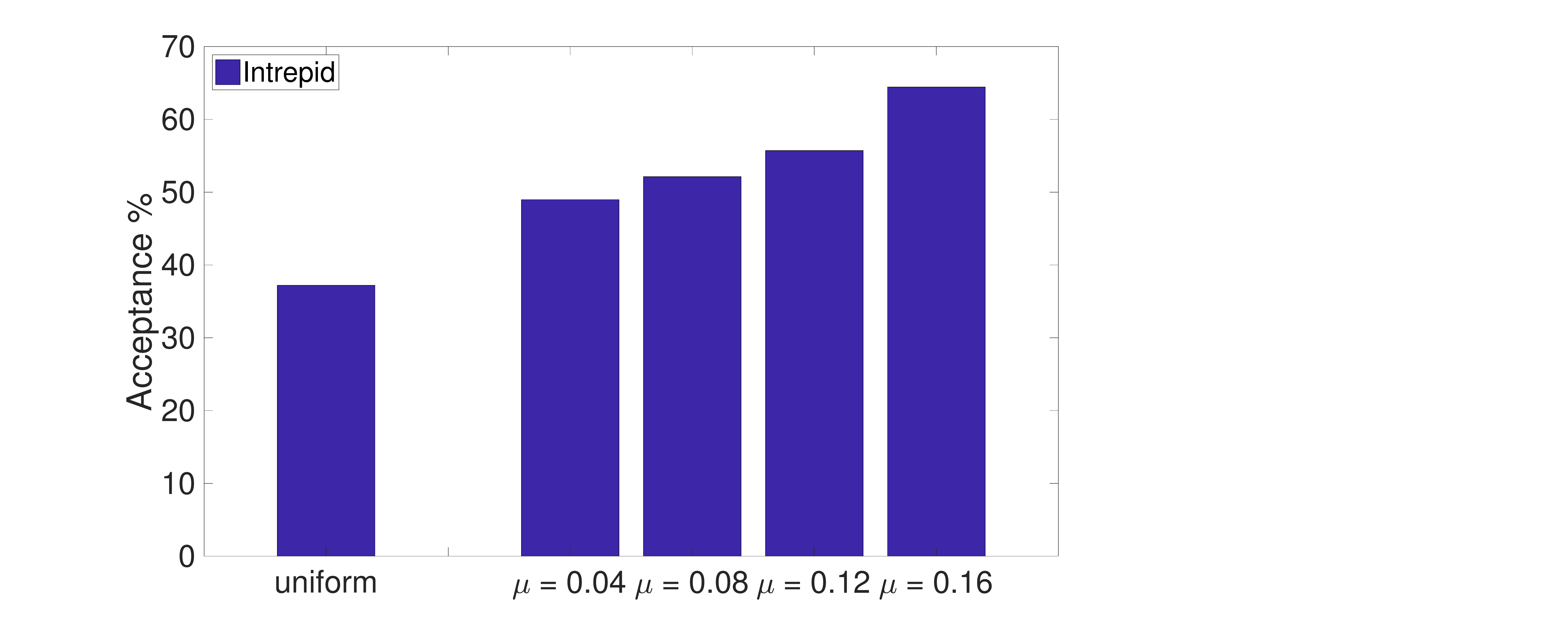}
  \caption{Effects of user flexibility models on acceptance.}\label{fig:acceptance_models}
\endminipage\hfill
\minipage{0.32\textwidth}%
  \includegraphics[width=0.85\linewidth]{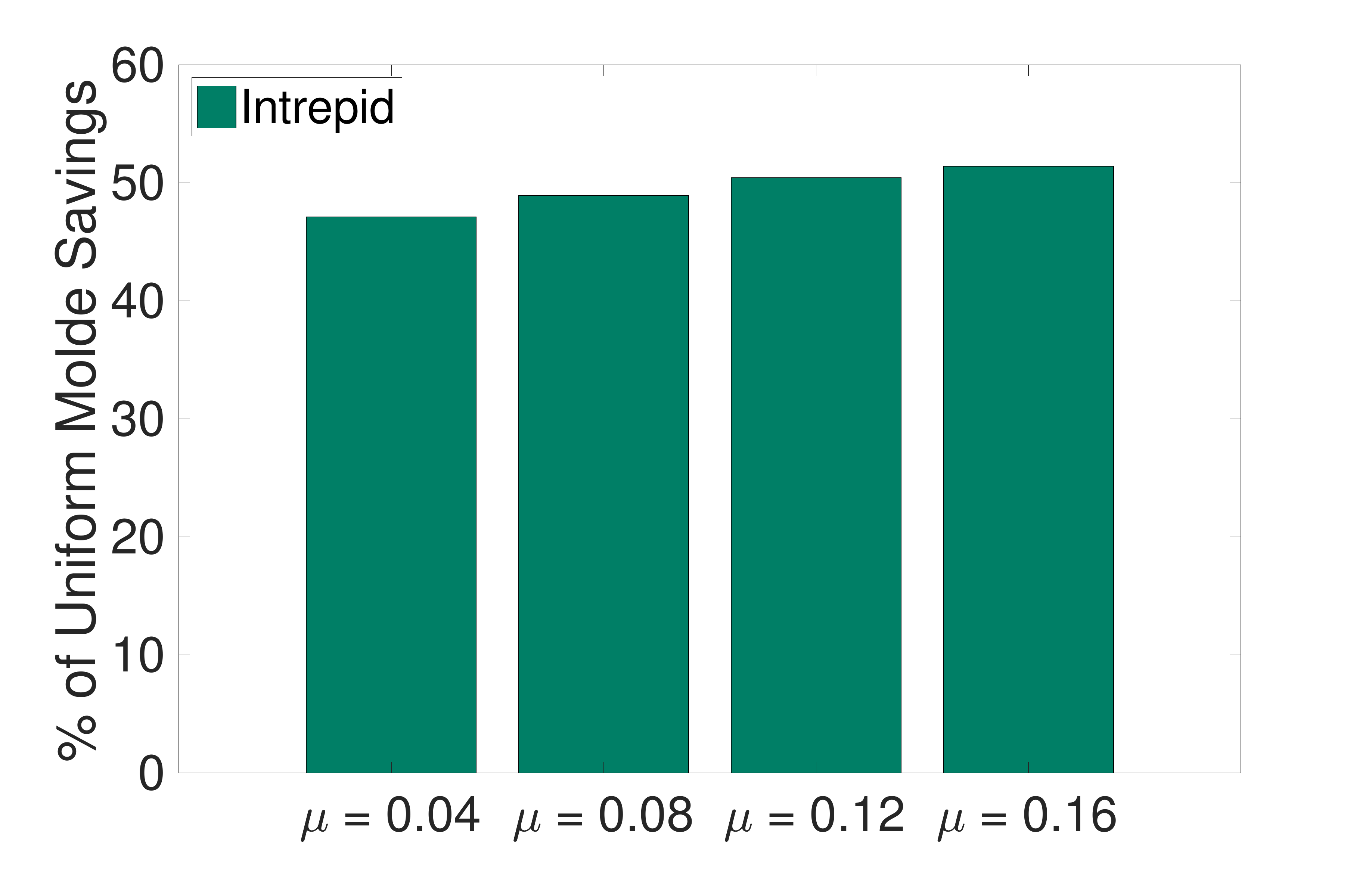}
  \caption{Effects of exponential models on the absolute financial savings.}\label{fig:savings_models}
\endminipage
\end{figure*}

Figure \ref{fig:gain_summary} shows a summary of the device operation scheduling results on both the Intrepid and Genetx datasets. The figure shows a positive utility, in terms of decrease in energy price, for the proposed device scheduling model.  
Specifically, for the Intrepid dataset with an acceptance rate of the proposed schedules of $64\%$, we can achieve a savings of $32\%$ and $38\%$ in spot and regulation markets, respectively. 
On the other hand, in the Genetx dataset we can achieve a higher acceptance of 79$\%$, leading to higher spot and regulation Market cost reduction of 42$\%$ an 39$\%$, respectively. 
This is due to the higher prediction accuracy that can be achieved in the (more regularized) synthetic dataset, as we further explain in Section \ref{sec:ev_of_predictions}.

To analyze the relationship between user acceptance and financial benefits, we show a comparison of two different user flexibility models on the scheduling objective function shown in Eq. \ref{eq:gain}. The first is a simple \textit{uniform} model that does not distinguish between different schedules, i.e. $\forall t, t_{es},t_{le} \quad P(A =a | t, t_{es}, t_{le}) = 1 $. Then, we apply our proposed user flexibility model with different learning rates $\mu$ for Eq. \ref{eq:gradient} that were found empirically, controlling how \textit{fast} does our flexibility model adapts to each new user feedback.

Fig. \ref{fig:acceptance_models} shows the effects of the different user flexibility models on the scheduling acceptance. The adaptive flexibility model obtains a higher acceptance, from $48\%$ for the learning rate $\mu = 0.04$, up to $64\%$ for the learning rate $\mu = 0.16$. The uniform model achieves the lowest acceptance of $36\%$. This comes at the cost of lower financial benefits. If we do not consider the user acceptance, i.e., every schedule is accepted, the uniform model obtains the highest amount of savings. Fig. \ref{fig:savings_models} shows the percentage of savings, compared to savings of the uniform model, for the adaptive flexibility models. However, the higher savings for the uniform model comes at the cost of lower user acceptance, which entails a larger risk of user response fatigue. On the other hand, the adaptive user-flexibility model trade-off the savings with the acceptability ensuring efficient implementation of the proposed DR scheme.

The results show that the more the flexibility model adheres to the user preference, the higher the acceptance. Moreover, although our flexibility model obtains overall lower financial benefits, a DR schema that does not take into account the user preference will ultimately fail at being accepted, in the long term denying all benefits to the involved parties. Ultimately, the trade-off between the user preference over financial benefits and flexibility can be learned online, when user feedback, in terms of their scheduling acceptance/rejection, is available.

\subsection{Evaluation of Flexoffer Prediction}
\label{sec:ev_of_predictions}
The effectiveness of the device operation scheduling highly depends on the quality of the predicted device activities and the generated \fo s. 
If the \fo\ models a device flexibility interval that is distant, in terms of time or duration, from the real flexibility interval, the chances the resulting schedule be rejected will increase. 
Hence, the acceptability of the proposed schedule and the associated utility is influenced by underlying prediction models. 

In Table \ref{tab:pred_results} we start by showing the results of the prediction phase.
Here we show the quality of the predictions using our proposed combination of day-level and hour-level prediction, defined as \textit{2-levels prediction}. 
Moreover, we also show the quality of the predictions obtained with a \textit{1-level prediction}, performed by applying only the hour-level prediction (with linear regression) and selecting the 10$\%$ activations with the associated highest probability, yielding approximately the same number of predicted activations as the 2-levels prediction.  
In both datasets the 2-levels model outperforms the simple 1-level model, leading to higher accuracy in classifying which days will present an activation and lower error in the prediction of the hour of activation.
Additionally, in the Genetx dataset, we obtained an overall better prediction quality that leads to the higher scheduling results shown in Figure \ref{fig:gain_summary}, confirming a correlation between predictions and scheduling results.

\begin{table}
\centering
\caption{Prediction measures for day-level prediction (Day Accuracy) and hour-level prediction (RMSE).}
\label{tab:pred_results}
\begin{tabular}{|c|c|c|c|c|}
\hline
\multirow{2}{*}{} & \multicolumn{2}{c|}{2-levels Prediction} & \multicolumn{2}{c|}{1-level Prediction} \\ \cline{2-5} 
                  & Day Acc.         & Hour RMSE        & Day Acc.         & Hour RMSE        \\ \hline
Intrepid          & 66\%                 & 4.1              & 48\%                 & 6.88             \\ \hline
Genetx            & 80\%                 & 3.36             & 54\%                 & 6.78             \\ \hline
\end{tabular}
\end{table}

\begin{figure*}
\minipage{0.32\textwidth}
  \includegraphics[width=0.9\linewidth]{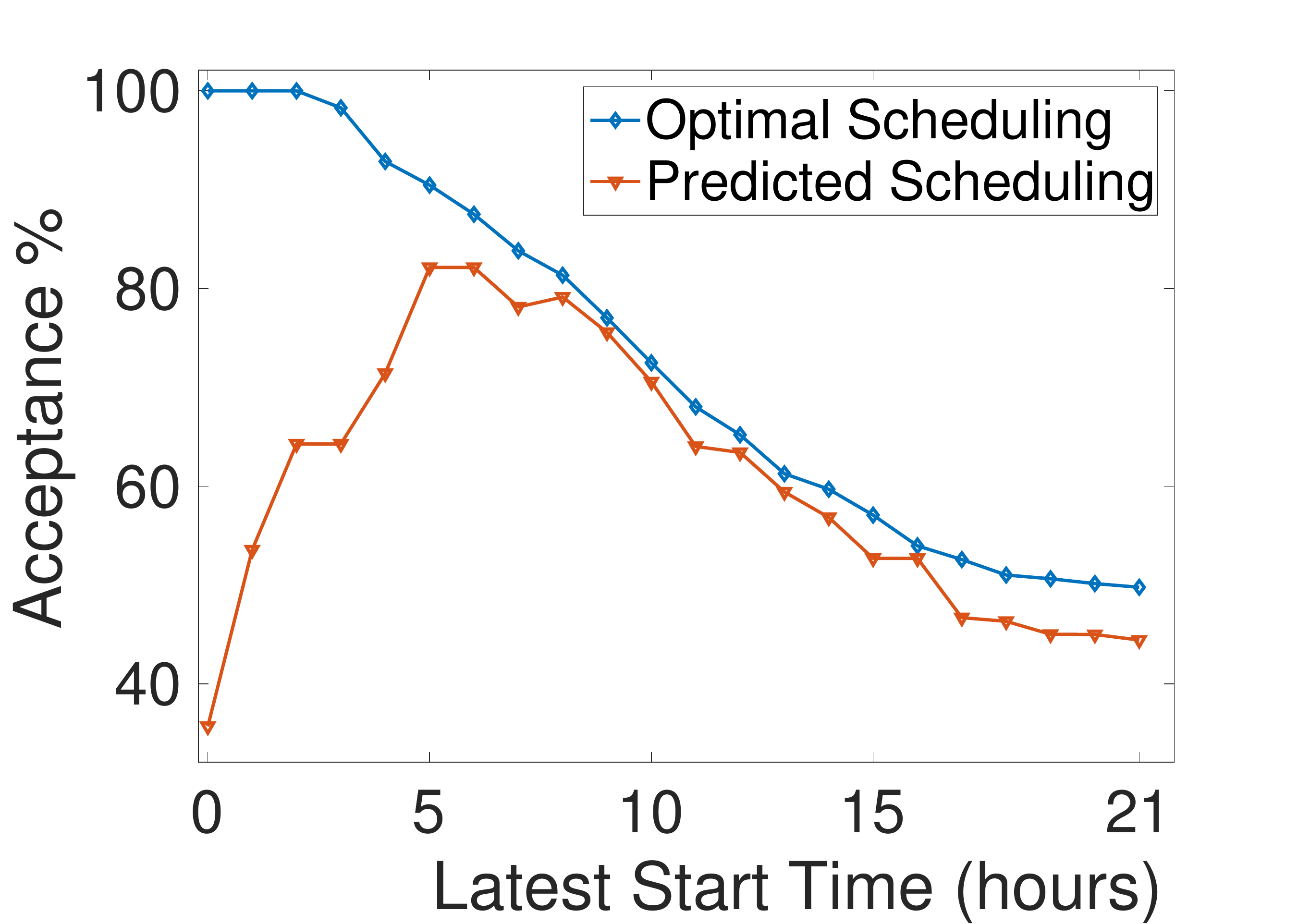}
  \caption{Discounted effect of predictions on acceptance.}\label{fig:flex_accept}
\endminipage\hfill
\minipage{0.32\textwidth}
  \includegraphics[width=0.9\linewidth]{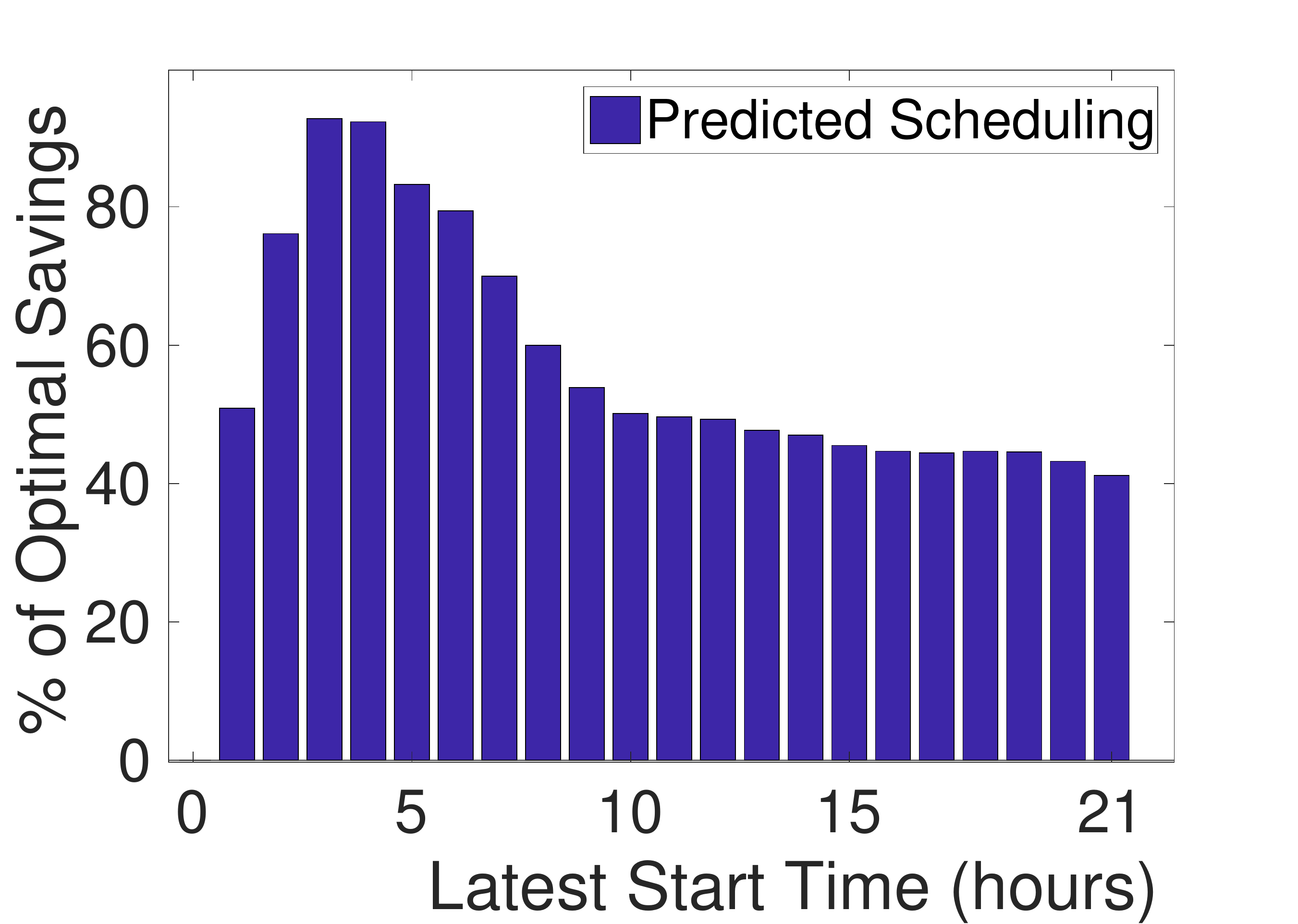}
  \caption{Discounted effect of predictions on savings.}\label{fig:flex_eval}
\endminipage\hfill
\minipage{0.32\textwidth}%
  \includegraphics[width=0.85\linewidth]{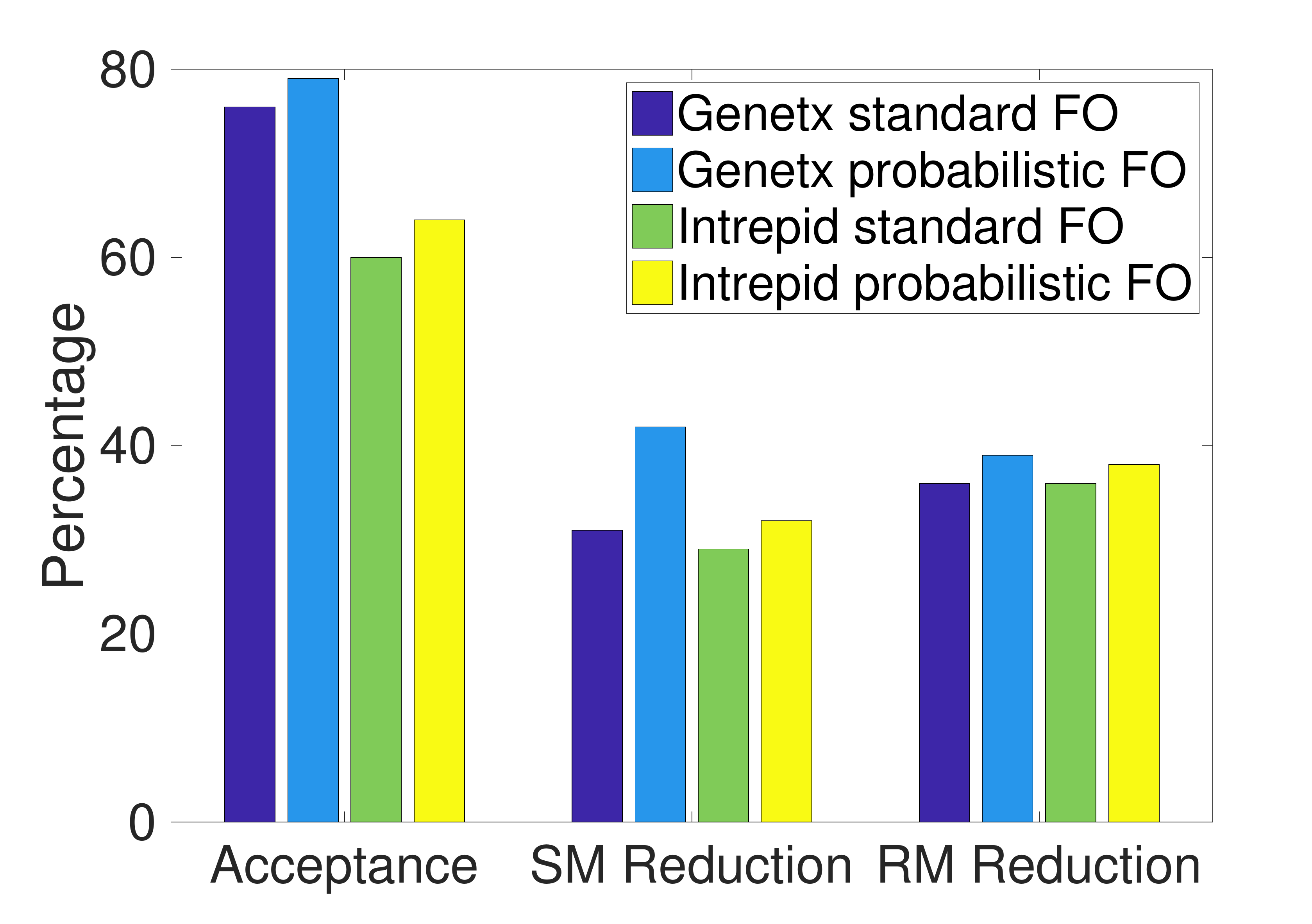}
  \caption{Comparison of standard \fo s and probabilistic \fo s.}\label{fig:std_vs_prob}
\endminipage
\end{figure*}

To evaluate the exact effect of the prediction models on the final scheduling results, we compare both financial savings and acceptance in two different scenarios. The first is the \textit{predicted scenario}, where day and hour of the earliest start time of a \fo\ is predicted by our prediction models. The second is the \textit{ideal scenario}, where we assume we assume the start time of a \fo\ is already known. In both scenarios, we introduce time flexibility by manually setting the latest start time of each \fo\ between 0 and 22 hours after the earliest start time. Again, in our experiments, we have used two simple forecasting models, Naive Bayesian and Linear regression model for day-level and hour-level predictions, respectively. We want to emphasize that short-term load forecasting is a largely researched field, and work on highly accurate forecasting model is outside of the scope of this paper.

Figure \ref{fig:flex_accept}, compares the user acceptability of the prescribed schedule for ideal and predicted scenarios at various time flexibility. The figure demonstrates that at a lower time-flexibility the user acceptability for the \textit{predicted scenario} is very low compared to the \textit{ideal scenario}. However, for a time-flexibility of more than 4 hours, the user acceptability are almost comparable for both the scenarios. 

This behavior is mainly due to the fact that at lower time-flexibilities the scheduling is not able to rectify the prediction error resulting in the lower acceptability. For e.g.,  a prediction error of -3 hours can be rectified if the device has $\geq$ 3 hours of time-flexibility, but not with $<$ 3 hours of time-flexibility. Therefore, we can see a significant improvement in user acceptability with the increase in time-flexibility. However, after a certain threshold  (5 hours in our case) the acceptability gradually decreases because of the larger flexibility a scheduler might overshoot the actual flexibility duration. For example, with 10 hours of time flexibility and +5 hours of prediction error, there is a high chance that a scheduler schedule a flexible demand after the user preferred timestamp of the next $ready$ action, resulting in the rejection of the proposed schedule. 

Similarly, Figure \ref{fig:flex_eval} illustrates the percentage of the ideal savings that can be achieved by the \textit{predicted scenario}. From the figure, we can see that even with the lower time flexibilities the \textit{predicted scenario} can achieve up to 90\% of the savings of the \textit{ideal scenario}. However, for higher time-flexibilities savings is at a range of 40\% of the \textit{ideal scenario}. As before, this is mainly because for higher time-flexibility, user acceptability decreases reducing the savings. 

Finally, Figure \ref{fig:std_vs_prob} shows the scheduling results, evaluated on the Genetx dataset, of using our probabilistic \fo\ compared to standard a \fo\ model. As it can be seen, modeling flexibility uncertainty improves the scheduling effectiveness by increasing the user acceptance, and therefore higher the financial benefits. In an uncertain scenario, standard \fo s miss useful device flexibility, leading to the lower probability of scheduling acceptance.

\subsection{Discussion}

First, we showed how our approach can provide users with effective device operation schedules and how it can reduce the energy cost of at least $30\%$ in both Spot and Regulation market, while maintaining a scheduling acceptance of at least $60\%$ on a real dataset.

Second, we demonstrated the importance of trade-off between quality of service, scheduling acceptance, and financial savings in order to successfully deploy a DR scheme. 
Our flexibility estimation model can increase user acceptance from less than $40\%$ up to $60\%$ while keeping a financial saving that is approximately $50\%$ of the one where no user comfort is taken into account, improving the customization of the DR scheme to specific users.

Third, the high stochasticity in the user behavior can lead to incorrect predictions of the user energy needs. However,  generic prediction models can lead to scheduling acceptance and financial savings that are close to the optimal, when the user provides sufficient flexibility. The higher prediction and scheduling results obtained in the synthetic dataset Genetx hint to a correlation between prediction and scheduling quality that allow an even higher margin of improvement if more sophisticated prediction models are used. In this highly stochastic scenario, our proposed probabilistic \fo\ for modeling flexibility under uncertainty can bring an improvement, with regards to the scheduling effectiveness, over traditional flexibility models, such as standard \fo s. 

\section{Conclusions}
\label{sec:conclusions}

We presented a novel DR scheme for user-oriented direct load-control of residential appliance operations. Instead of relying on user surveys and demanding interaction to evaluate the user utility, we proposed a data-driven approach for estimating user utility functions, purely based on available load consumption data. Moreover, we presented an online technique for learning the user utility functions, which adaptively model the users'   preference over time. 
Furthermore, our scheme is based on a scheduling scheme that transparently prescribes the users with optimal device operation schedules that take into account both financial benefits and user comfort, in order to reduce the threat of user response fatigue.  

The experimental results, performed on two different datasets, consistently showed that our approach can successfully reduce energy costs while preserving user comfort.
On a real dataset, our approach yields savings of approximately $32\%$ in the Spot Market, and $38\%$ in the Regulation Market, with $64\%$ of user acceptance of the proposed schedules.
Further, the results demonstrated that the proposed adaptive learning model could capture user flexibility with an acceptable accuracy generating positive utility to all involved players. 
Additionally, even under the stochasticity, the predicted scenario was able to achieve $90\%$ of the savings in the ideal scenario. 
Finally, our proposed probabilistic \fo\  for modeling flexibility under uncertainty can bring an improvement, with regards to the scheduling effectiveness, over traditional flexibility models.

Future work will establish statistical models to evaluate the scheduling of multiple devices and households. Also, the generic models applied in this paper, although already successful, allow for a margin of improvement. 
In this regard, we will explore more robust prediction techniques for device-level load forecasting by exploiting the information of multiple residential devices (e.g., kitchen appliances, HVAC, lights, etc.) to analyze patterns the user behavior, in order to improve both the prediction results and the estimation of the user flexibility. 
Additionally, we will investigate the use of flexibility-oriented prediction error measures that align the prediction of device activity to the scheduling of flexible demand.

\begin{acks}
This work was supported in part by the GoFLEX project funded under the Horizon 2020 program.
\end{acks}


\end{document}